\documentclass[twocolumn, pra]{revtex4}
\usepackage{graphicx}
\usepackage{dcolumn}
\usepackage{bm}
\usepackage{textcomp}
\usepackage{amssymb}
\usepackage{mathrsfs}
\usepackage{amsmath}
\allowdisplaybreaks[4]
\usepackage[colorlinks, linkcolor=blue, anchorcolor=blue,citecolor=blue]{hyperref}

\begin{document}
\title{Optomechanical Microwave Quantum Illumination in Weak Coupling Regime}

\author{Wen-Juan Yang$ ^{1, 2}$, Xiang-Bin Wang$ ^{1, 2, 3\footnote{Email
Address: xbwang@mail.tsinghua.edu.cn}}$}
\affiliation{ \centerline{$^{1}$State Key Laboratory of Low
Dimensional Quantum Physics, Tsinghua University, Beijing 100084,
People's Republic of China}\centerline{$^{2}$ CAS Center for Excellence and Synergetic Innovation Center in Quantum Information and Quantum Physics,}
 \centerline{University of Science and Technology of China, Hefei, Anhui 230026, China}\centerline{$^{3}$ Jinan Institute of Quantum Technology, Jinan, Shandong 250101, China}}

\begin{abstract}
We propose to realize microwave quantum illumination in weak coupling regime based on multimode optomechanical systems. In our proposal the multimode together with a frequency-mismatch process could reduce mechanical thermal noise. Therefore, we achieve a significant reduction of error probability than conventional detector in weak coupling regime. Moreover, we optimize the signal-to-noise ratio for limited bandwidth by tuning the delay time of entangled wave-packets.
\end{abstract}

\maketitle
\parskip=0 pt

\section{Introduction}
Quantum entanglement is a key ingredient in quantum information processing\cite{e1, e2, e3}. In practice, entanglement can be easily destroyed by environment noise\cite{envir1, envir2}. Quantum illumination(QI) can benefit from entanglement in target detection even it is under these entanglement-destroying noise\cite{e3, qi1, qi2, qi3, qi4, qi5, qi6}.

The aim of quantum illumination is to detect low-reflective target which is embedded in a bright background thermal bath. Half of a pair of entangled optical beams is sent out to interrogate the target region. Then the returned and the retained signal beam are used to decide the presence or absence of the object. Even though the fragile entanglement is easily destroyed by the bright thermal noise, quantum-illumination protocol still has remarkable advantage over classical probe protocol\cite{qi2, qi4, qi5, qi6}. Several experiments have realized the practical protocol in quantum sensing\cite{exp1, exp2, exp3}. Recently, Weedbrook et al.\cite{qi6} show that quantum discord exhibits the role of preserving the benefits of quantum illumination while entanglement is broken.

It is advantageous to operate the frequency of the signal which interrogates the object in microwave region\cite{qi4, qi5, exp1, exp2, exp3}. But so far there is no efficient way to detect single photon in microwave region while in the visible light frequency region, ultrasensitive detection of single-photon have been achieved. Such that the detection of microwave signals via the detection of their entangled optical signals is a more efficient way\cite{micro1, micro2, micro3, micro4}. Optomechanical systems could be a good candidate to realize the entanglement of microwave and optical fields by using mechanical motion\cite{oms1,oms2, oms3, oms4, oms5, oms6, oms7, oms8, oms9, oms10}. However, the existing proposal is supposed to work with rotating wave approximation(RWA) which requests weak many-photon couplings of optomechanical systems\cite{qi5}. In addition, it is extremely difficult to achieve strong coupling in optical regime for a system to generate entanglement between optical and microwave modes. Moreover, plenty of photons in cavities can induce bistability which may give rise to experimental difficulties\cite{oms9}. Therefore, it is meaningful to study the generation of entanglement in weak interaction regime. Also, the existing study is limited to the situation that the output signals are monochromatic. A study for the more practical case of finite bandwidth is needed.

In this paper, we propose a novel realization of quantum illumination in weak coupling regime based on multimode optomechanical systems. The transmitter is an optomechanical system consisting of a two-mode microwave cavity coupling with a two-mode optical cavity via a mechanical resonator. In this system, the output microwave signal and the optical signal of the two cavities are entangled. The receiver is a similar optomechanical system which converts the reflected microwave signal into optical signal. The retained optical mode of the transmitter and the optical mode of the receiver are then sent to the photon-detectors to make a joint measurement. We use a two-mode and off-resonant(with frequency detuning $\delta$) process to minimize the mechanical thermal noise. Consequently, our method can achieve a significant reduction of error probability than classical system of the same transmitted energy. Our method works in the weak many-photon coupling regime where rotating-wave approximation works well. Moreover, we optimize the delay time of the microwave signal's filter function. This makes it possible for our method to work with a finite bandwidth of signals. In this case, we can still achieve an improvement of signal-to-noise ratio by $48\%$ than that of classical detection given that signals-bandwidth $\sigma$ which equals cavities-bandwidth $\kappa$ .

The paper is organized as follows. In Sec. II we describe the system and derive the quantum Langevin equations for the Hamiltonian of our proposed system. In Sec. III we calculate the error probability of the proposed QI system, give an analytical expression for the delay time and analyze the limited bandwidth situation.  Finally, Sec. IV summarizes the main conclusions of this work.

\section{Multi-mode Optomechanical System}
\begin{figure}
\centering
\includegraphics[width=0.48\textwidth]{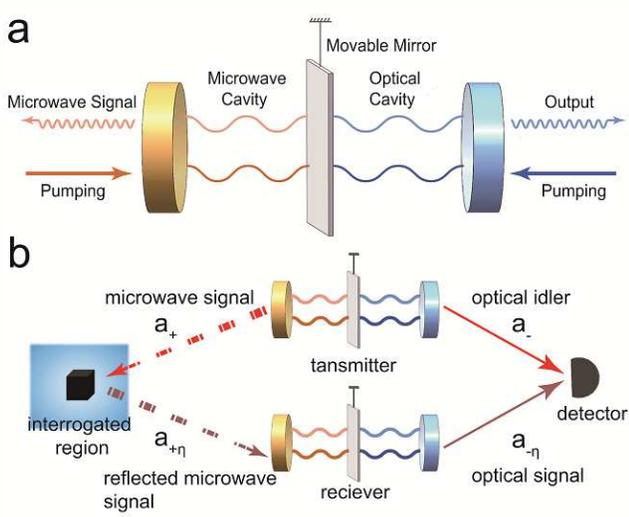}\\
\caption{(a) Optomechanical system consisting of a two-mode microwave cavity coupling with a two-mode optical cavity via a mechanical resonator. (b) Realization of  microwave quantum illumination in hybrid optomechanical system. The transmitter entangles microwave signal and optical signal. The receiver converts the reflected microwave signal into optical signal and performs a phase-conjugate operation.}\label{fig1}
\end{figure}
As illustrated in Fig. \ref{fig1} (a),
 we consider an optomechanical system including a single-mode mechanical resonator coupled to both a two-mode microwave cavity and a two-mode optical cavity. The two microwave modes are at frequencies $\omega_{+,p}$, $\omega_{+}$ and two optical modes are at frequencies $\omega_{-,p}$, $\omega_{-}$ respectively. The mechanical mode is at frequency $\omega_b$. Cavity modes $\hat{a}_{+,p}$ and $\hat{a}_{-,p}$ are resonantly driven by lasers. The Hamiltonian of the system is\cite{micro4, model1, model2, model3}
\begin{eqnarray}
\hat{H}&=&\sum_{j=\pm}(\omega_{j,p}\hat{a}_{j,p}^\dagger\hat{a}_{j,p}+\omega_{j}\hat{a}_{j}^\dagger\hat{a}_{j})+\omega_M\hat{b}^\dagger\hat{b}\nonumber\\
&&-\sum_{j=\pm}jg_j(\hat{a}_{j,p}^\dagger+\hat{a}_{j}^\dagger)(\hat{a}_{j,p}+\hat{a}_{j})(\hat{b}+\hat{b}^\dagger)\nonumber\\
&&+\sum_{j=\pm}\Omega_j(\hat{a}_{j,p}e^{i\omega_{j,p}t+i\phi_{j}}+\mathrm{H.c.}), \label{ham}
\end{eqnarray}
where $\omega_{\pm}=\omega_{\pm,p}\pm J$, $\Omega_j$ and $\phi_j$ are the intensities and phases of the driving lasers respectively. Given the frequency mismatch fulfilling $-\omega_M\ll\delta=\omega_M-J\ll\omega_M$, under the resolved-sideband approximation and rotated-wave approximation, the Hamiltonian can be linearized as\cite{model2, model3}
\begin{eqnarray}
\hat{H}_{1}=\delta\hat{b}^\dagger\hat{b}-\frac{G_1}{2}(\hat{a}^\dagger_{+}\hat{b}+\hat{a}_{+}\hat{b}^\dagger)
-\frac{G_2}{2}(\hat{a}_{-}\hat{b}+\hat{a}^\dagger_{-}\hat{b}^\dagger), \label{ham1}
\end{eqnarray}
where $G_{1}/2=g_{+}\alpha_{+}$, $G_2/2=-g_{-}\alpha_{-}$, $\alpha_{\pm}=\frac{-i\Omega_{\pm} e^{-i\phi_{\pm}}}{\kappa/2}$ and $\kappa$ is the cavity decay rate. Coupling terms $G_1$ and $G_2$ can be chosen to be real and positive by adjusting the laser phases $\phi_{+}$ and $\phi_{-}$ respectively. The Hamiltonian shows that the emitted photons from modes $\hat{a}_{+}$ and $\hat{a}_{-}$ could be entangled via intermediate phonon mode $\hat{b}$. Taking cavity dissipation $\kappa$ and mechanical dissipation $\gamma$ into consideration, the Langevin equations of Eq. (\ref{ham1}) are\cite{inout}
\begin{eqnarray}
\dot{\hat{a}}_{+}&=&i\frac{G_1}{2}\hat{b}-\frac{\kappa}{2}\hat{a}_{+}-\sqrt{\kappa}\hat{a}_{+,in},\nonumber \\
\dot{\hat{a}}_{-}&=&i\frac{G_2}{2}\hat{b}^\dagger-\frac{\kappa}{2}\hat{a}_{-}-\sqrt{\kappa}\hat{a}_{-,in},\\
\dot{\hat{b}}&=&i\frac{G_1}{2}\hat{a}_{+}+i\frac{G_2}{2}\hat{a}_{-}^\dagger-(i\delta+\frac{\gamma}{2})\hat{b}-\sqrt{\gamma}\hat{b}_{in}.\nonumber
\end{eqnarray}
The equations can be solved after Fourier transformations where $\hat{O}[\omega]=\int_{-\infty}^{\infty}\hat{O}(t)e^{i\omega t}dt$(with $\hat{O}=\hat{a}_{+},\hat{a}_{-},\hat{b}$). Applying the input-output relations $\hat{O}_{out}=\sqrt{\Gamma}\hat{O}+\hat{O}_{in}$(with $\Gamma=\kappa, \gamma$)\cite{inout}, we obtain
\begin{eqnarray}
\begin{aligned}
\hat{a}_{+}[\omega]=&A_{+}\hat{a}_{+,in}[\omega]-B\hat{a}_{-,in}^\dagger[-\omega]+C_{+}\hat{b}_{in}[\omega],\\
\hat{a}_{-}[-\omega]=&B^{*}\hat{a}_{+,in}^\dagger[\omega]+A_{-}\hat{a}_{-,in}[-\omega]+C_{-}\hat{b}_{in}^\dagger[\omega].
\end{aligned}
\end{eqnarray}
The correlation functions of input mode satisfy
\begin{eqnarray}
\begin{aligned}
\langle\hat{a}^\dagger_{+,in}(t)\hat{a}_{+,in}(t^\prime)\rangle&=&n_{+}^{T}\delta(t-t^\prime),\\
\langle\hat{a}^\dagger_{-,in}(t)\hat{a}_{-,in}(t^\prime)\rangle&=&n_{-}^{T}\delta(t-t^\prime),\\
\langle\hat{b}^\dagger_{in}(t)\hat{b}_{in}(t^\prime)\rangle&=&n_{b}^{T}\delta(t-t^\prime),
\end{aligned}
\end{eqnarray}
where $n_{+}^{T}$, $n_{-}^{T}$ and $n_{b}^{T}$ are thermal photon numbers of the two cavity modes and phonon thermal number of mechanical mode respectively. We obtain
\begin{widetext}
\begin{subequations}
\begin{align}
n_{+}[\omega,\omega^\prime]&=\langle\hat{a}_{+,out}^\dagger[\omega]\hat{a}_{+,out}[\omega^\prime]\rangle
=2\pi[4n_{b}^{T}\frac{G_1^2\gamma\kappa}{|G_1^2-G_2^2+[\gamma +2 i (\delta -\omega )] (\kappa -2 i \omega )|^2}\nonumber\\&+4\frac{(G_1G_2\kappa)^2 }{|[G_1^2-G_2^2+(\gamma +2 i (\delta -\omega )) (\kappa -2 i \omega )](\kappa -2 i \omega )|^2}]\delta(\omega-\omega^\prime),\\
n_{-}[-\omega,-\omega^\prime]&=\langle\hat{a}_{-,out}^\dagger[-\omega]\hat{a}_{-,out}[-\omega^\prime]\rangle
=2\pi[4(1+n_{b}^{T})\frac{G_2^2\gamma\kappa}{|G_1^2-G_2^2+[\gamma +2 i (\delta -\omega )] (\kappa -2 i \omega )|^2}\nonumber\\&+4\frac{(G_1G_2\kappa)^2 }{|[G_1^2-G_2^2+(\gamma +2 i (\delta -\omega )) (\kappa -2 i \omega )](\kappa -2 i \omega )|^2}]\delta(\omega-\omega^\prime),\\
x[\omega,-\omega^\prime]&=\langle\hat{a}_{+,out}[\omega]\hat{a}_{-,out}[-\omega^\prime]\rangle=2\pi\delta(\omega-\omega^\prime)
[-2G_1G_2\kappa\nonumber\\
&\frac{G_1^2(\kappa-2i\omega )+(\kappa +2i\omega)(G_2^2-(\gamma +2i(\delta -\omega)) (\kappa -2i\omega))+2 (1+n_{b}^{T})\gamma(\kappa ^2+4\omega ^2)}{(G_1^2-G_2^2+(\gamma +2 i (\delta -\omega )) (\kappa -2 i \omega ))(G_1^2-G_2^2+(\gamma -2 i(\delta -\omega) ) (\kappa +2 i \omega ))(\kappa ^2+4 \omega ^2)}].
\end{align}
\end{subequations}
\end{widetext}
Here we assume that $n_{+}^{T}=n_{-}^{T}=0$. In practice, in order to minimize the error probability of output signals, an increase of repeat rate is needed. Therefore, we need to consider output signals within certain frequency bandwidth. After projecting output signals onto wave-packet modes, $\hat{a}_{\pm,f}=\int_{-\infty}^{\infty}f^{*}_{\pm}(t)\hat{a}_{\pm, out}(t)dt$. Here $f_{\pm}(t)$ are filter functions which satisfy $
\int^{\infty}_{-\infty}|f_{\pm}(t)|^2dt=1$.

We use logarithmic negativity to measure the entanglement of two cavity modes. The logarithmic negativity is given by\cite{enlog}
\begin{eqnarray}
E_{N}=\mathrm{max}(0,-\log_22\eta^{-}),
\end{eqnarray}
where $\eta^{-}$ is the smallest partial transposed symplectic eigenvalue of matrix $V$. The components of the covariance matrix have the form $V_{ij}=\frac{1}{2}\langle \xi_i\xi_j+\xi_j\xi_i\rangle$ where $\xi=[\hat{x}_{+}, \hat{p}_{+}, \hat{x}_{-}, \hat{p}_{-}]^{T}$ with $\hat{x}_{\pm}=(\hat{a}_{\pm,f}+\hat{a}_{\pm,f}^\dagger)/\sqrt{2}$ and $\hat{p}_{\pm}=(\hat{a}_{\pm,f}-\hat{a}_{\pm,f}^\dagger)/(i\sqrt{2})$. Thus we have
\begin{eqnarray}
\textbf{V}=\left(
\begin{array}{cccc}
V_{11}+\frac{1}{2}& 0 & V_{13} & V_{14}\\
0 & V_{11}+\frac{1}{2} & V_{14} & -V_{13}\\
V_{13} & V_{14} & V_{33}+\frac{1}{2} & 0\\
V_{14} & -V_{13} & 0 & V_{33}+\frac{1}{2}\\
\end{array}
\right),
\end{eqnarray}
where
\begin{eqnarray}
V_{11}&=&\langle\hat{a}_{+,f}^\dagger\hat{a}_{+,f}\rangle=\int_{-\infty}^{\infty}|f_{+}[\omega]|^2n_{+}[\omega]d\omega,\nonumber\\
V_{33}&=&\langle\hat{a}_{-,f}^\dagger\hat{a}_{-,f}\rangle=\int_{-\infty}^{\infty}|f_{-}[\omega]|^2n_{-}[\omega]d\omega,\nonumber\\
V_{13}&=&\mathrm{Re}(\langle\hat{a}_{+,f}\hat{a}_{-,f}\rangle)=\mathrm{Re}(\int_{-\infty}^{\infty}f_{+}^{*}[\omega]f_{-}^{*}[-\omega]x[\omega]d\omega),\nonumber\\
V_{14}&=&\mathrm{Im}(\langle\hat{a}_{+,f}\hat{a}_{-,f}\rangle)=\mathrm{Im}(\int_{-\infty}^{\infty}f_{+}^{*}[\omega]f_{-}^{*}[-\omega]x[\omega]d\omega),\nonumber
\end{eqnarray}
with
\begin{eqnarray}
n_{+}[\omega]=\frac{1}{2\pi}\int_{-\infty}^{\infty}n_{+}[\omega,\omega^\prime]d\omega^\prime,\nonumber\\
n_{-}[\omega]=\frac{1}{2\pi}\int_{-\infty}^{\infty}n_{-}[\omega,\omega^\prime]d\omega^\prime,\nonumber\\
x[\omega]=\frac{1}{2\pi}\int_{-\infty}^{\infty}x[\omega,-\omega^\prime]d\omega^\prime.\nonumber
\end{eqnarray}
\begin{figure}
  \centering
  \includegraphics[width=0.45\textwidth]{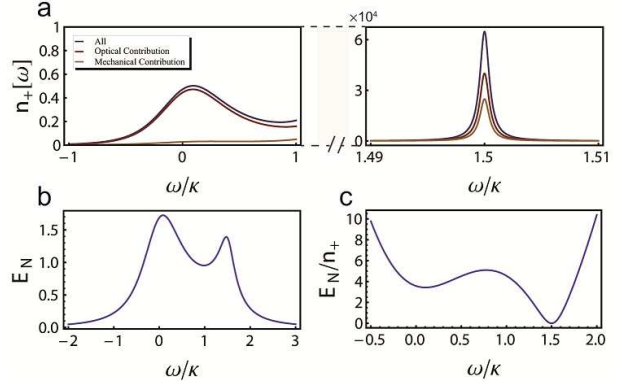}
  \caption{(a) The spectrum of photon number $n_{+}[\omega]$ around $\omega=0$ and around $\omega=\delta$ where $G/\kappa = 1.0$, $\gamma/\kappa= 0.001$, $\delta/\kappa = 1.5$, $n_{b}^T =61.945$($\omega_{+}/2\pi=10\mathrm{GHz}$, $T_b=30\mathrm{mK}$), $n_{+}^{T} = 0$, $n_{-}^{T} = 0$. (b) The spectrum of logarithmic-negativity $E_{N}$. (c) The spectrum of logarithmic-negativity over output photon number $E_{N}/n_{+}$.}\label{en}
\end{figure}
We draw the spectrum of $n_{+}[\omega]$, $E_{N}[\omega]$ and $E_{N}[\omega]/n_{+}[\omega]$ in Fig. \ref{en}. As shown in Fig. \ref{en}(a), output photon number $n_{+}[\omega]$ has two peaks. One is at $\omega\approx0$ with width $\sim\kappa$ and the other is at $\omega\approx\delta$ with width $\sim\Gamma$. When $\delta$ approaches $0$, the two peaks emerge together. In this case, $n_{+}[\omega=0]\gg1$ and the maximum entanglement appears at $\omega=0$, and the ratio($E_N/n_{+}$) of the entanglement to the photon number goes to the minimum. This is the reason why the optimum ratio $E_N/n_{+}$ in the previous literature Ref. \cite{qi5} appears at $G_1\gg G_2$ but not at the optimum entanglement point where parameter $G_1\approx G_2$. In our case where $\delta\neq 0$, even though the ratio $E_N/n_{+}$ is the second minimum valley at around $\omega=0$, it still have a considerable value.
\section{Quantum illumination}
The input-output relation for the receiver is
\begin{eqnarray}
\hat{a}_{-\eta}[-\omega]=B^{*}\hat{a}_{+R,in}^{\dagger}[\omega]+A_{-}\hat{a}_{-,in}^\prime[-\omega]+C_{-}\hat{b}_{in}^{\prime\dagger}[\omega],
\end{eqnarray}
where $\hat{a}_{+R,in}$ is the signal reflected from interrogated target region. We have $\hat{a}_{+R,in}=\hat{a}_{B}$ when there is no low reflective object in the target region(hypothesis $H_0$) and $\hat{a}_{+R,in}=\sqrt{\eta}\hat{a}_{+}+\sqrt{1-\eta}\hat{a}_{B}$  when there contains low reflective object in the target region(hypothesis $H_1$)\cite{qi4}. Here $\hat{a}_B$ is the background thermal mode in the interrogated region. The mean photon number is $n_{B}$ or $n_{B}/(1-\eta)$ providing the absence or presence of low reflective object.

We measure the photon numbers of mixed modes $\hat{c}_{\eta,\pm}=(\hat{a}_{-\eta}\pm\hat{a}_{-})/\sqrt{2}$. The object's absence or presence is determined by the photon number difference which is
$\hat{N}_c=\sum_{k=1}^{M}(\hat{N}_{c,+}^{(k)}-\hat{N}_{c,-}^{(k)})$
where $\hat{N}_{c,\pm}^{(k)}=\hat{c}_{\eta,\pm}^{(k)\dagger}\hat{c}_{\eta,\pm}^{(k)}$ and $M=t_mW_m$ is the independent,
identically distributed (iid) signal-idler-mode pairs. Thus the error probability of receiver is given by\cite{qi4}
\begin{eqnarray}
P_{\mathrm{QI}}^{M}=\frac{\mathrm{erfc}(\sqrt{\mathrm{SNR}_{\mathrm{QI}}^{M}/8})}{2},
\end{eqnarray}
where the signal-to-noise ratio satisfies
\begin{widetext}
\begin{eqnarray}
\mathrm{SNR}_{\mathrm{QI}}^{M}=\frac{4M[(\langle\hat{N}_{c,+}\rangle_{H_1}-\langle\hat{N}_{c,-}\rangle_{H_1})
-(\langle\hat{N}_{c,+}\rangle_{H_0}-\langle\hat{N}_{c,-}\rangle_{H_0})]^2}
{(\sqrt{(\langle\Delta\hat{N}_{c,+}-\Delta\hat{N}_{c,-}\rangle_{H_0})^2}+\sqrt{(\langle\Delta\hat{N}_{c,+}
-\Delta\hat{N}_{c,-}\rangle_{H_1})^2})^2},\label{snr}
\end{eqnarray}
\end{widetext}
where
\begin{widetext}
\begin{subequations}
\begin{align}
\langle\hat{N}_{c,+}\rangle_{H_0}-\langle\hat{N}_{c,-}\rangle_{H_0}&=0,\\
\langle\hat{N}_{c,+}\rangle_{H_1}-\langle\hat{N}_{c,-}\rangle_{H_1}&=2\sqrt{\eta}\mathrm{Re}[\langle B\hat{a}_{+,f}\hat{a}_{-,f}\rangle],\label{12b}\\
(\langle\Delta\hat{N}_{c,+}-\Delta\hat{N}_{c,-}\rangle_{H_0})^2&=\langle\hat{N}_{c,+}\rangle_{H_0}(\langle\hat{N}_{c,+}\rangle_{H_0} + 1)+\langle\hat{N}_{c,-}\rangle_{H_0}(\langle\hat{N}_{c,-}\rangle_{H_0} + 1) - \frac{(\langle\hat{a}_{-\eta,f}^\dagger\hat{a}_{-\eta,f}\rangle_{H_0} -\langle\hat{a}_{-,f}^\dagger\hat{a}_{-,f}\rangle)^2}{2},\\
(\langle\Delta\hat{N}_{c,+}-\Delta\hat{N}_{c,-}\rangle_{H_1})^2&=\langle\hat{N}_{c,+}\rangle_{H_1}(\langle\hat{N}_{c,+}\rangle_{H_1} + 1)+\langle\hat{N}_{c,-}\rangle_{H_1}(\langle\hat{N}_{c,-}\rangle_{H_1} + 1) - \frac{(\langle\hat{a}_{-\eta,f}^\dagger\hat{a}_{-\eta,f}\rangle_{H_1} -\langle\hat{a}_{-,f}^\dagger\hat{a}_{-,f}\rangle)^2}{2}\nonumber\\
&-2(\sqrt{\eta}\mathrm{Im}[\langle B\hat{a}_{+,f}\hat{a}_{-,f}\rangle])^2.
\end{align}
\end{subequations}
\end{widetext}

\begin{figure}
  \centering
  \includegraphics[width=0.5\textwidth]{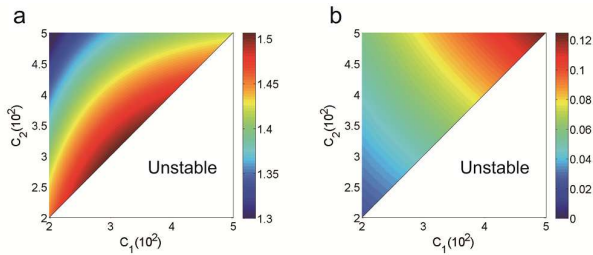}\\
  \caption{(a) QI advantage over classical system. $F$ versus $C_1$ and $C_2$. Parameter settings are $\gamma/\kappa= 0.001$, $\delta/\kappa = 1.5$, $n_{b}^T =61.945$, $n_{+}^{T} = 0$, $n_{-}^{T} = 0$, $\eta=0.07$ and $n_{B} = 610$(room temperature $T=273\mathrm{K}$). (b) Photon number $n_{+}[\omega=0]$ versus cooperative parameters $C_1$ and $C_2$. For $C_1, C_2\gg 1$ and $\kappa\gg\gamma$ the stability condition could be $C_1\geq C_2$\cite{oms10}.}\label{p}
\end{figure}

In Fig. \ref{p} (a), we plot $F\equiv\mathrm{SNR}_{\mathrm{QI}}^{M}/\mathrm{SNR}_{\mathrm{coh}}^{M}$ with cooperative parameters $C_1=\frac{G_1^2}{\kappa\gamma}$ and $C_2=\frac{G_2^2}{\kappa\gamma}$ in resonant frequency $\omega=0$. Here $\mathrm{SNR}_{\mathrm{coh}}^{M}$ is the signal-to-noise ratio of classical microwave transmitter which is equivalent to $4\eta MV_{11}/(2n_B+1)$\cite{qi5}. From the figure, we can see that the maximum $F$ appears at $C_1\approx C_2$ and a relative large $F$ could be obtained with weak many-photon optomechanical coupling where $G_1/2, G_2/2\ll\kappa$. Photon number $n_{+}[0]$ is shown in fig. \ref{p}(b) with $C_1$ and $C_2$.

\begin{figure}
  \centering
  \includegraphics[width=0.35\textwidth]{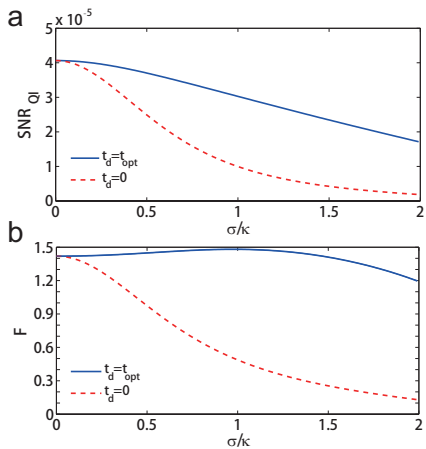}\\
  \caption{(a) Signal-to-noise ratio $\mathrm{SNR}_{\mathrm{QI}}$ versus bandwidth $\sigma/\kappa$. Parameter settings are $C_1=500$, $C_2=500$, $\gamma/\kappa=0.001$, $n_{+}^T=0$, $n_{-}^T=0$, $n_b^T =61.945$, $nB = 610$, $\delta/\kappa=1.5$, $\eta=0.07$. (b) $F$ versus bandwidth $\sigma/\kappa$.}\label{width}
\end{figure}

Let us now discuss the situation where the output signals have a bandwidth $\sigma$, i.e., filter functions $f_{+}[\omega]$ and $f_{-}[\omega]$ have a bandwidth $\sigma$. The filter functions have the form $f_{+}[\omega]=e^{i\omega t_d}/\sqrt{\sigma}$, $f_{-}[\omega]=1/\sqrt{\sigma}$ for $\omega\in[-\frac{\sigma}{2},\frac{\sigma}{2}]$ and $f_{\pm}[\omega]=0$ for $\omega\in(-\infty,-\frac{\sigma}{2})$ or $(\frac{\sigma}{2},\infty)$.  Here $f_{+}[\omega]$ has a time-delay term $e^{i\omega t_d}$ as discussed in Ref. \cite{model2}. To find optimal $t_{opt}$ at which the signal-to-noise ratio decreases slowly with the bandwidth, we revisit Eq. (\ref{12b}). We have\cite{model2}
\begin{eqnarray}
\langle B\hat{a}_{+,f}\hat{a}_{-,f}\rangle&=&\frac{1}{\sigma}\int_{-\frac{\sigma}{2}}^{\frac{\sigma}{2}}
e^{-i\omega t_d}x'[\omega]=\frac{1}{\sigma}\int_{-\frac{\sigma}{2}}^{\frac{\sigma}{2}}\nonumber\\&&[
e^{-i\omega t_d}|x'[\omega]|e^{i(\varphi(0)+\varphi^{\prime}(0)\omega+...)}],
\end{eqnarray}
where $x'[\omega]=[A_{+}|B|^2+C_{+}C_{-}B(n_b^T+1)]$ and $\varphi$ is the phase of $x'[\omega]$. The elimination of integrand's phase dependence on $\omega$ would improve the value of signal-to-noise ratio. If bandwidth $\sigma$ is small, then $t_d=\varphi^{\prime}(0)=\frac{d\varphi(\omega)}{d\omega}|_{\omega=0}$ could be the optimal delay time which reads
\begin{eqnarray}
t_{opt}&=&\frac{1}{\gamma}[\frac{2}{1+4(\frac{\delta}{\gamma})^2}+4\frac{\gamma}{\kappa}\nonumber\\&&+\frac{n_b^T+\frac{1}{2}+C}{
  (n_b^T+\frac{1}{2}+C)^2+(\frac{\delta}{\gamma})^2}],
\end{eqnarray}
where we assume that $C_1=C_2=C$ and $n_{+}^T=n_{-}^T=0$. In Fig. \ref{width} (a), we show $\mathrm{SNR}_{\mathrm{QI}}$ with bandwidth $\sigma$ for delay time $t_d=0$ and $t_d=t_{opt}$. It can be easily seen that the signal-to-noise ratio indeed decreases slowly with $\sigma$ increases for $t_d=t_{opt}$ compared to that for $t_d=0$. Figure of merit $F$ with $\sigma$ is shown in Fig. \ref{width} (b). The advantage of QI over classical system decreases rapidly for $t_d=0$ while QI always has advantage over classical ones even for relative large $\sigma$ for $t_d=t_{opt}$. The performance of QI is about $48\%$ better than classical system at $\sigma=\kappa$.

\begin{figure}
  \centering
  \includegraphics[width=0.35\textwidth]{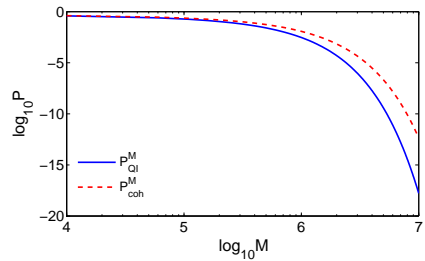}\\
  \caption{Error probability $P_{\mathrm{QI}}^{M}$ and $P_{\mathrm{coh}}^{M}$ versus $\log_{10}M$ for $\sigma/\kappa=1$. See Fig. \ref{width} for the other parameter values. }\label{errorp}
\end{figure}

Fig. \ref{errorp}
plots error probability $P_{\mathrm{QI}}^{M}$ and $P_{\mathrm{coh}}^{M}$ versus $\log_{10}M$ for $\sigma=\kappa$. Microwave(optical) photon number $\bar{n}_{+}\approx 0.09$($\bar{n}_{-}\approx 0.09$). QI has orders of magnitude better performance in error probability than classical system.

\section{Conclusion}
In summary, we have proposed a scheme to achieve optomechanical QI in weak coupling regime. The frequency mismatch $\delta$ minimizes the mechanical thermal noise and increases the ratio $E_N/n_{+}$. As a result, we achieve a substantial reduction of error probability than classical system at $C_1=C_2$ compared to that of Ref.\cite{qi5} at $C_1\gg C_2$. Consequently, our quantum illumination proposal works in weak coupling regime. Furthermore, we optimize the delay time of microwave signal's wave packets. Finally, we find a $48\%$ improvement of signal-to-noise ratio for $\sigma=\kappa$.

\section*{ACKNOWLEDGMENTS}
 We acknowledge the financial support in part by the $10000$-Plan of Shandong province (Taishan Scholars), NSFC grant No. $11174177$ and $60725416$, Open Research Fund Program of the State Key Laboratory of Low-Dimensional Quantum Physics Grant No. $\mathrm{KF}201513$; and the key $\mathrm{R}\&\mathrm{D}$ Plan Project of Shandong Province, grant No. $2015\mathrm{GGX}101035$.

\end{document}